\documentclass[conference]{IEEEtran}
\usepackage{graphicx}

\usepackage{numprint}
\usepackage{textcomp}
\usepackage{booktabs}
\usepackage{amsmath}
\usepackage{amssymb}
\usepackage{textcomp}
\usepackage{dirtytalk}
\usepackage{float}
\usepackage{xcolor}
\usepackage{verbatim}
\usepackage{multirow}

\ifCLASSINFOpdf
\else
\fi
\begin{document}
%
\title{Towards Perspective-Based Specification of Machine Learning-Enabled Systems}


\author{\IEEEauthorblockN{Hugo Villamizar, Marcos Kalinowski, and Hélio Lopes}
\IEEEauthorblockA{Informatics Department\\
Pontifical Catholic University of Rio de Janeiro\\
Rio de Janeiro, Brazil\\
\{hvillamizar, kalinowski, lopes\}@inf.puc-rio.br}
}


%


\maketitle

\begin{abstract}

Machine learning (ML) teams often work on a project just to realize the performance of the model is not good enough. Indeed, the success of ML-enabled systems involves aligning data with business problems, translating them into ML tasks, experimenting with algorithms, evaluating models, capturing data from users, among others. Literature has shown that ML-enabled systems are rarely built based on precise specifications for such concerns, leading ML teams to become misaligned due to incorrect assumptions, which may affect the quality of such systems and overall project success. In order to help addressing this issue, this paper describes our work towards a perspective-based approach for specifying ML-enabled systems. The approach involves analyzing a set of 45 ML concerns grouped into five perspectives: objectives, user experience, infrastructure, model, and data. The main contribution of this paper is to provide two new artifacts that can be used to help specifying ML-enabled systems: (i) the perspective-based ML task and concern diagram and (ii) the perspective-based ML specification template.

\end{abstract}


%
\IEEEpeerreviewmaketitle

\section{Introduction} 
\label{sec:introduction}

Companies from all sectors are increasingly incorporating machine learning (ML) components into the software that supports their operations. We call this ML-enabled systems. The main difference between ML-enabled systems and non-ML is that data, to some extent, replaces code when determining the behavior of ML-enabled systems. This is challenging from the point of view of software engineering (SE). For instance, data should be tested and models be validated just as thoroughly as code, but there is a lack of best practices on how to do so~\cite{arpteg2018software}.

Data scientists often work on a project just to realize that the performance of the model is not good enough. However, it is important to understand several aspects. For example, where will the model run? What data will it have access to? How fast does it need to be? What is the business impact of a false positive? A false negative? How should the model be tuned to maximize business results? In short, a model is just one component of a system as a whole. 

Within the SE discipline, requirements engineering (RE) has drawn the attention of researchers and practitioners to the fact that ML can benefit from this perspective~\cite{dalpiaz2020requirements}. Literature has shown that current research on the intersection between RE and ML mainly focuses on using ML techniques to support RE activities rather than on exploring how RE can improve the development of ML-based systems~\cite{ahmad2021s}. In addition, most ML models lack requirements specifications since current RE practices are not well defined and organized for ML~\cite{ishikawa2019engineers}. This indicates that there is an incredible amount of work to be done between the development of a model, the incorporation of it into a system and the eventual sustainable customer impact.

Given this landscape, we propose a perspective-based approach for specifying ML-enabled systems. The approach relies on a set of concerns grouped into five perspectives: objectives, user experience, infrastructure, model, and data. The approach, built based on the literature~\cite{villamizar2021requirements}, industrial experiences working on ML-enabled systems~\cite{kalinowski2020lean}, and an initial industrial validation of relevant concerns \cite{villamizar2022catalogue}, covers relevant perspectives and practical concerns that have not been considered as part of related work (e.g.,~\cite{chuprina2021towards, nakamichi2020requirements, nalchigar2021modeling}).
\section{Perspective-based Specification of ML Systems} 
\label{sec:approach}


The proposed perspective-based approach provides two artifacts to support the specification of ML-enabled systems: (i) a perspective-based ML task and concern diagram to be analyzed; and (ii) a specification template to be filled. These artifacts were designed based on findings from a literature review~\cite{villamizar2021requirements}, industrial experiences~\cite{kalinowski2020lean}, and an initial validation of the ML concerns~\cite{villamizar2022catalogue}. While the literature review showed us how ML could benefit from the RE perspective, what particular properties and research opportunities could be addressed, our industrial experience allowed us to understand how ML works in practice from early to final stages. Hence, we established links between theory and practice to learn more about the context in which our approach operates.

\subsection{Perspective-Based ML Task and Concern Diagram} 
\label{subsec:ml_diagram}

\begin{figure*}[ht]
    \centering
    \includegraphics[width=0.8\textwidth]{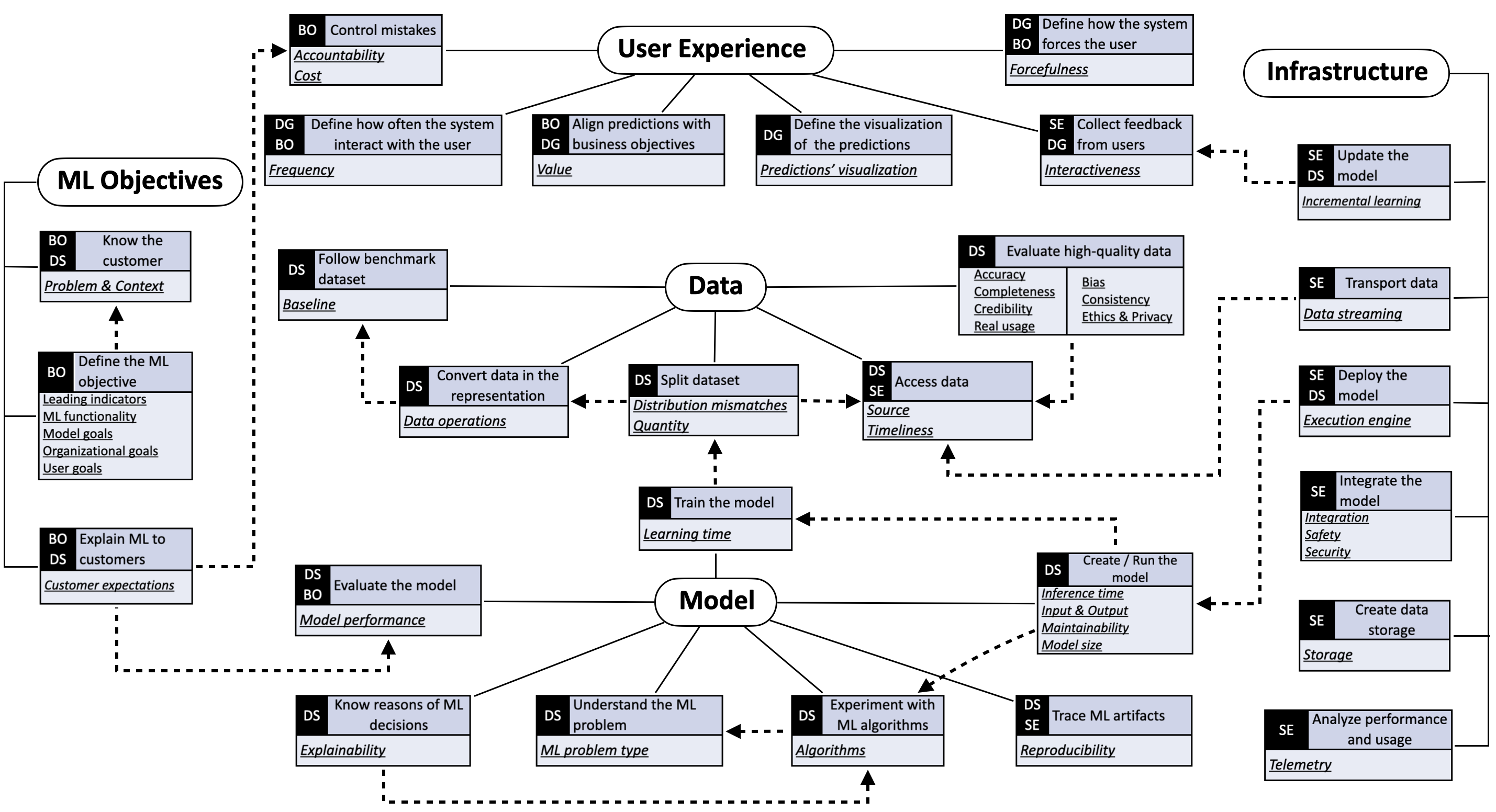}
    \caption{Perspective-based ML task and concern diagram}
    \label{fig:method_overview}
\end{figure*}

The perspective-based ML task and concern diagram shown in Figure~\ref{fig:method_overview} consists of five perspectives: objectives, user experience, infrastructure, model, and data. Together these perspectives mediate the involvement of actors such as business owners (BO), designers (DG), software engineers (SE), and data scientists (DS). Each perspective contains a set of concerns that should be analyzed by a requirement engineer with the support of at least one actor. In a previous study~\cite{villamizar2022catalogue}, we conducted a focus group session with eight software professionals with experience developing ML-enabled systems to validate the importance, quality and feasibility of using the concerns.

The diagram contains five rounded rectangles that represent the ML perspectives. Each perspective is divided into rectangles that connect a task (at the top right) to one or more concerns (at the bottom). Note that each task-concern has at least one actor suggested (at the top left) related to the execution of the task and the analysis of the concerns. For instance, typically in ML projects, DS are tasked for training, running, and evaluating models. These tasks involve implicit concerns that are not easily identified such as learning time, inference time, and model size (Cf. model perspective in Figure~\ref{fig:method_overview}). The diagram also models dashed arrows that represent dependencies between the tasks. It is noteworthy that dependencies typically implicitly imply an influence in the opposite direction. For example, to transport data, that is a task of the infrastructure perspective that can require implementing real time analytics, it is necessary first to access the data and define the source. In the following, we describe the role of the actors involved in our approach. 

\textbf{BO} typically understand how to plan ML projects in order to make good decisions. How to connect business objectives with ML outcomes? What represents future success from a business point of view? \textbf{SE} typically understand how the system will interact with the model DS produce and identify the components that need to exist within a successful ML-enabled system. What are the pros and cons of running a model in a client or in a service? \textbf{DS} typically understand the constraints regarding the ML models. For instance, what quality properties the model should consider? What domain restrictions may apply? How fast does it need to be? \textbf{DG} typically understand how to improve the experience of end-users of the system and its ML inferences. For analyzing some tasks and concerns associated to the DG, if possible, end-users should also be involved. For instance, where and how often should the ML outcomes appear? How forcefully should they appear?

Note that specific concerns can benefit from involving more than one actor in the analysis. E.g., the execution engine concern involves knowledge in charge of DS and may also involve expertise of SE that provide the ML-enabled system infrastructure. Hereafter, we further describe the perspectives and concerns of the diagram.

\subsubsection{ML Objectives Perspective} 
\label{subsubsec:goals_measures_artifact}

Bridging the gap between the high-level goals and the detailed properties of ML models is one of the most common causes of failure to succeed~\cite{barash2019bridging}. Table~\ref{table:goal_properties} presents the concerns related to ML objectives that could influence the specification of ML-enabled systems.


\begin{table}[ht]
\scriptsize
\centering
\caption{ML objectives perspective}
\label{table:goal_properties}
\begin{tabular}{|p{2cm}|p{5.8cm}|}
\hline
\multicolumn{1}{|c|}{\textbf{Concern}} & \multicolumn{1}{c|}{\textbf{Description (Addressing this concerns involves ... )}}\\ \hline

Customer expectations      & 
Specifying expectations of customers and end-user in terms of how the system should behave (e.g., how often they expect predictions to be right or wrong). \\ \hline

Leading indicators       & 
Specifying measures correlating with future success, from the business' perspective. This could include the users' affective states when using the ML-enabled system (e.g., customer sentiment and engagement).  \\ \hline

ML functionality        & 
Specifying the ML results in terms of functionality that the model will provide (e.g., classify customers, predict probabilities). \\ \hline

Model goals       & 
Specifying metrics and acceptable measures the model should achieve (e.g., for classification problems this could involve accuracy $\geq$ X\%, precision $\geq$ Y\%, recall $\geq$ Z\%).  \\ \hline

Organizational goals     & 
Specifying measurable benefits ML is expected to bring to the organization. E.g., increase the revenue in X\%, increase the number of units sold in Y\%, number of trees saved. \\ \hline

Problem \& Context                 &  
Specifying the problem that ML will address and its context before coding. ML must be targeted at the right problem.\\ \hline

User goals               & 
Specifying what the users want to achieve by using ML. E.g., for recommendation systems this could involve helping users finding content they will enjoy.\\ \hline

\end{tabular}
\end{table}


\subsubsection{User Experience (UX) Perspective} 
\label{subsubsec:user_exp_perspective}

Better ML includes building better experiences of using ML. Connecting the predictions with users is critical for achieving success. However, this requires deep analysis. For instance, one mistake in the way of interacting with the user, and all the work spent on data pre-processing and modeling may be wasted. The goal of this perspective is to create effective UX so that users can interact appropriately and understand what is going on with the predictions of the model. Table~\ref{table:user_exp_properties} shows the UX concerns and a brief description of each one.


\begin{table}[ht]
\scriptsize
\centering
\caption{UX Perspective}
\label{table:user_exp_properties}
\begin{tabular}{|p{1.8cm}|p{6cm}|}
\hline
\multicolumn{1}{|c|}{\textbf{Concern}} & \multicolumn{1}{c|}{\textbf{Description (Addressing this concerns involves ... )}}\\ \hline

Accountability  & Specifying who is responsible for unexpected model results or actions taken based on unexpected model results. \\ \hline


Cost            & Specifying costs involved in executing the inferences and also the user impact of a wrong model prediction. \\ \hline


Forcefulness    & Specifying how strongly the system forces the user to do what the model indicates they should (e.g., automatic or assisted actions).\\ \hline


Frequency       &  Specifying how often the system interacts with users (e.g., interact whenever the user asks for it or whenever the system thinks the user will respond). \\ \hline


Interactiveness    & Specifying what interactions the users will have with the ML-enabled system, (e.g., to provide new data for learning, or human-in-the-loop systems where models require human interaction). \\ \hline


Value           & Specifying the added value as perceived by users from the predictions to their work. \\ \hline


Prediction Visualization    & Specifying how the ML outcomes will be presented so that users can understand them (e.g., specifying dashboard and visualization prototypes for validation). \\ \hline

\end{tabular}
\end{table}


\subsubsection{Infrastructure Perspective} 
\label{subsubsec:infrastructure_perspective}

ML models need to be integrated with other services. This includes components such as ingesting and learning from new data. The adoption of ML is growing, but proper implementations are needed to fulfill their promise. Hence, software engineers play an important role to orchestrate these ML components. Table~\ref{table:infra_properties} presents the infrastructure concerns.


\begin{table}[hb]
\scriptsize
\centering
\caption{Infrastructure Perspective.}
\label{table:infra_properties}
\begin{tabular}{|p{1.5cm}|p{6.3cm}|}
\hline
\multicolumn{1}{|c|}{\textbf{Concern}} & \multicolumn{1}{c|}{\textbf{Description (Addressing this concerns involves ... )}}\\ \hline

Data streaming          & Specifying what data steaming strategy will be used (e.g., real time data transportation or in batches). \\ \hline


Execution engine        & Specifying how the model of the ML-enabled system will be executed and consumed (e.g., client-side, back-end, cloud-based, web service end-point). \\ \hline


Incremental learning    & Specifying the need for ML-enabled system abilities to continuously learn from new data, extending the existing model’s knowledge. \\ \hline


Integration             & Specifying the integration that the model will have with the rest of the system functionality.  \\ \hline


Safety       & Specifying how the system deals with risks to prevent dangerous failures. Critical systems that incorporate ML should analyze the probability of the occurrence of harm and its severity.\\\hline


Security     & Specifying how the system deals with security issues (e.g., vulnerabilities) to protect the data. ML systems often contain sensitive data that should be protected. \\ \hline


Storage                 & Specifying where the ML artifacts (e.g., models, data, scripts) will be stored. \\ \hline


Telemetry               & Specifying what ML-enabled system data needs to be collected. Telemetry involves collecting data such as clicks on particular buttons and could involve other usage and performance monitoring data.  \\ \hline

\end{tabular}
\end{table}


\subsubsection{Model Perspective} 
\label{subsubsec:model_perspective}

Building a model implies not only training an algorithm with data to predict or classify well some phenomenon. Many other aspects determine its success. The aim of this perspective is to provide a set of model concerns a requirements engineer should analyze. Table~\ref{table:model_properties} presents these concerns.


\begin{table}[ht]
\scriptsize
\centering
\caption{Model Perspective.}
\label{table:model_properties}
\begin{tabular}{|p{1.7cm}|p{6cm}|}
\hline
\multicolumn{1}{|c|}{\textbf{Concern}} & \multicolumn{1}{c|}{\textbf{Description (Addressing this concerns involves ... )}}\\ \hline

Algorithms     & Specifying the set of algorithms that could be used/investigated, based on the ML problem and other concerns to be considered (e.g., constraints regarding explainability or model performance, for instance, can limit the solution options). \\ \hline


Explainability       & Specifying the need to understand reasons of the model inferences. The model might need to be able to summarize the reasons of its decisions. Other related concerns, such as transparency and interpretability, may apply. \\ \hline


Inference time       & Specifying the acceptable time to execute the model and return the predictions.\\ \hline


Input \& Output       & Specifying the expected inputs (features) and outcomes of the model. Of course, the set of meaningful inputs can be refined/improved during pre-processing activities, such as feature selection. \\ \hline


Learning time        & Specifying the acceptable time to train the model.\\ \hline


Maintainability   & Specifying the need for preparing the model to go through changes with reasonable effort (e.g., refactoring, documentation, automated redeployment). \\ \hline


ML problem type & Specifying the problem type tackled by the ML algorithm (e.g., classification, regression, clustering, extract information from text).\\ \hline


Model performance  & 
Specifying the metrics used to evaluate the model (e.g., precision, recall, F1-score, mean square error) and measurable performance expectations. \\ \hline


Model size           & Specifying the size of the model in terms of storage and its complexity (e.g., for decision trees there might be needs for pruning).\\ \hline


Reproducibility      & Specifying the need for replicating the model creation process and its experiments. \\ \hline

\end{tabular}
\end{table}


\subsubsection{Data Perspective} 
\label{subsubsec:data_perspective}

Data is essential for ML-enabled systems. Poor data will result in inaccurate predictions, which is referred to in the ML context as “garbage in, garbage out”. Hence, ML requires high-quality input data. From the viewpoint of RE, it is clear that data constitutes a new type of requirements~\cite{challa2020faulty},~\cite{vogelsang2019requirements}. Based on the Data Quality model defined in the standard ISO/IEC 25012~\cite{iso25012}, we elaborate on the data perspective. Table~\ref{table:data_properties} presents the data concerns.


\begin{table}[ht]
\scriptsize
\centering
\caption{Data Perspective.}
\label{table:data_properties}
\begin{tabular}{|p{1.6cm}|p{6.2cm}|}
\hline
\multicolumn{1}{|c|}{\textbf{Concern}} & \multicolumn{1}{c|}{\textbf{Description (Addressing this concerns involves ... )}}\\ \hline

Accuracy         & Specifying the need to get correct data.\\ \hline


Baseline       & Specifying the need for a baseline dataset approved by a domain expert that reflects the problem. It is employed to monitor other data acquired afterwards. \\ \hline


Bias             & Specifying the need to get data fair samples and representative distributions. \\ \hline


Completeness     & Specifying the need to get data containing sufficient observations of all situations where the model will operate.\\ \hline


Consistency      & Specifying the need to get consistent data in a specific context.\\ \hline


Credibility      & Specifying the need to get true data that is believable and understandable by users.\\ \hline


Data operations      & Specifying what operations must be applied on the data (e.g., data cleaning and labeling). \\ \hline


Distribution mismatches    & Specifying expected data distributions and how data will be split into training and testing data.\\ \hline


Ethic \& privacy     & Specifying the need to get data to prevent adversely impacting society (e.g., listing potential adverse impacts to be avoided). \\ \hline


Quantity         & Specifying the expected amount of data according to the type of the problem and the complexity of the algorithm. \\ \hline


Real usage       & Specifying the need to get real data representing the real problem.\\ \hline


Source           & Specifying from where the data will be obtained. \\ \hline


Timeliness       & Specifying the time between when data is expected and when it is readily available for use.\\ \hline


\end{tabular}
\end{table}



\subsection{Perspective-based ML Specification Template} 
\label{subsec:ml_document}

Requirements specification produces a description of part of what should be built and delivers it for approval and requirements management. Based on the perspective-based ML task and concern diagram, the requirements engineer can identify, with the support of the actors, which tasks and concerns should be analyzed and described as part of the specification. 

The division into perspectives, the influence and dependency relationships between tasks and concerns and the suggested involvement of actors can help in this analysis. For instance, it can be seen in Figure \ref{fig:method_overview} that if the solution requires to know reasons of ML decisions (task) then the explainability concern arises. Explainability depends on the chosen algorithm, and should be specified considering this aspect. For classification problems, decisions taken by neural networks are typically less explainable than decisions taken by decision trees. 

To support the specification, we provide an ML requirement specification template that allows specifying the applicable ML-enabled system concerns. Inherently to the nature of ML-enabled system projects, some types of requirements (e.g., required algorithm, required data pre-processing operations, model performance requirements) are uncertain at the beginning of the project, mainly due to a common need of experimentation with data and algorithms to get a better understanding on achievable requirements. Hence, they may be refined or more precisely specified as the project progresses. The ML specification template highlights these concerns with the letter \say{E}. Our online material~\footnote{https://doi.org/10.5281/zenodo.6503232} presents an example of the ML-enabled system specification template, filled out retroactively with the support of practitioners with information of two real industrial ML-enabled system projects.

\section{Concluding Remarks} 
\label{sec:concluding_remarks}

The success of ML-enabled systems involves understanding business context and problems, translating them into ML tasks, designing and experimenting with algorithms, evaluating models, implementing telemetry, among other tasks. This needs to be considered from early stages of ML software development. Based on a literature review~\cite{villamizar2021requirements}, practical experiences~\cite{kalinowski2020lean}, and an initial
validation of the ML specification concerns\cite{villamizar2022catalogue}, we presented our work towards an approach for specifying ML-enabled systems considering five different perspectives: objectives, UX, infrastructure, model, and data. The main contribution of this paper is to provide two new artifacts that can be used to help specifying ML-enabled systems: (i) the perspective-based ML task and concern diagram and (ii) the ML requirements specification template. We believe that these artifacts can help requirements engineers with an overview of ML concerns that should be analyzed together with business owners, designers, software engineers, and data scientists to specify ML-enabled systems. Future work includes exploring how traditional and new RE techniques can use the perspective-based ML task and concern diagram and the ML specification template. In addition, we plan to conduct empirical studies to evaluate the completeness and feasibility of using the artifacts presented in this work.   

\section*{Acknowledgment} 
We would like to thank the CAPES agency for financial support. 

\bibliographystyle{IEEEtranS}
\bibliography{bibTex/sigproc} 

\end{document}